# High-speed, High-Resolution, Three-Dimensional Imaging of Threading Dislocations in β-$Ga_2O_3$ via Phase-Contrast Microscopy

Yukari Ishiakwa,[1,a)] Daiki Katsube,[1] Yongzhao Yao,[1,2] Koji Sato,[1] and Kohei Sasaki[3]

[1]Japan Fine Ceramics Center, 2-4-1, Mutsuno, Atsuta, Nagoya, 456-8587, Japan

[2]Mie University, 1577, Kurimamachiyacho, Tsu, 514-8507, Japan

[3]Novel Crystal Technology Inc., Hirosedai, Sayama, 350-1328, Japan

[a)]Author to whom correspondence should be addressed: yukari@jfcc.or.jp

This study presents a nondestructive, high-resolution method for three-dimensional imaging of threading dislocations inβ-$Ga_2O_3$ (010) using phase-contrast microscopy (PCM). A one-to-one correspondence between dislocation contrasts in PCM images and synchrotron X-ray topography (SR-XRT) images confirm the detection capability of PCM. Compared to SR-XRT, PCM provides enhanced spatial resolution, enabling the distinction of closely spaced dislocations with sub-10 μm separation. PCM facilitates direct visualization of dislocation propagation paths along the depth (z) direction by systematically shifting the focal plane into the crystal. In addition, the projection of stacked PCM images enables in-plane (XY) tracing of dislocation lines, providing insight into the preferred slip systems in β-$Ga_2O_3$. This work establishes PCM as a versatile and laboratory-accessible technique for three-dimensional, nondestructive characterization of dislocations across entire wide-bandgap semiconductor wafers within a practically acceptable time frame.

## I.INTRODUCTION

β-$Ga_2O_3$ is a promising ultrawide-bandgap semiconductor material with substantial potential for next-generation power electronics due to its high breakdown field and the availability of bulk single crystals via melt growth.[1–3] Despite these advantages, dislocations in β-$Ga_2O_3$ single crystals remain critical defects that impair device performance and reliability. Gaining insight into the three-dimensional propagation and slip behavior of dislocations is vital for enhancing crystal quality and device yield. To ensure device reliability, full-wafer inspection of all regions using laboratory-scale equipment is required. The inspection must be completed within a practically acceptable time frame for production efficiency—for example, within approximately one hour for a 6-inch wafer at the laboratory scale. However, nondestructive and high-resolution techniques capable of three-dimensional characterization of threading dislocations (TDs) across entire wafer areas within a short time remain limited.

X-ray topography (XRT) remains the only nondestructive technique capable of detecting dislocations across entire β-$Ga_2O_3$ wafer areas within a practically acceptable time frame (4–15 hrs/4inch wafer).[4–10] However, accurately determining the dislocation positions on the sample is frequently hindered by image deformation resulting from the non-perpendicular incidence of diffracted X-rays,[11] as well as distortions caused by domain boundaries.[8] In addition, in reflection geometry, XRT can detect dislocation structures only within approximately 10 μm from the surface. In contrast, transmission XRT employing the Borrmann effect[12,13] enables the detection of all dislocations in β-$Ga_2O_3$ as projection images. Techniques for reconstructing the three-dimensional structure of dislocations from such projection images acquired under various g-vectors have also been demonstrated.[14] However, reliably tracing dislocations within these projection images remains challenging.

Recently, dislocation detection in 4H-SiC using phase-contrast microscopy (PCM) —a classical optical microscopy technique invented by Frits Zernike in the 1930s—was reported by Hattori et al.[15,16] Since dislocation detection by PCM is attributed to variations in the refractive index induced by inelastic strain fields near dislocation cores,[16] similar detection is anticipated in β-$Ga_2O_3$. Furthermore, because a single PCM image of β-$Ga_2O_3$ with an area of 359 × 300 μm² can be acquired within 3 ms using PCM, capturing 170,000 images—corresponding to the area of a 6-inch wafer—would require only 510 s, indicating that PCM has the potential to enable full-wafer inspection within approximately one hour. Moreover, PCM can facilitate the observation of the three-dimensional (3D) structure of dislocations by adjusting the focal plane from the surface into the crystal. This study demonstrates the detection of dislocations in β-$Ga_2O_3$ using PCM, validated by a one-to-one correspondence between the contrasts in PCM images and those in synchrotron X-ray topography (SR-XRT) images of the same region. In addition, the three-dimensional structure of dislocations in β-$Ga_2O_3$ is visualized by shifting the focal plane.

## II. EXPERIMENTAL DETAILS

A 10 × 15 × 0.5 mm β-$Ga_2O_3$ (010) single crystal chip grown by edge-defined film-fed growth (EFG) was used.[17] Both surfaces of the chip were chemically and mechanically polished to prevent light scattering. The dislocation density of the chip was $8\times10^4$ $cm^{-2}$. SR-XRT images were obtained in reflection geometry (Supplementary Fig. 1) using the $\boldsymbol{g}=\bar{6}23$ ($\lambda$=0.1648nm, $\omega$=5°, $2\theta$=98.44°) diffraction vector, which was selected based on the discussions in Refs. 8 and 9. The images were recorded on Agfa Structurix D2 films. Measurements were performed at beamline BL3C of the High Energy Accelerator Research Organization.

Image deformation in the SR-XRT data was corrected via an affine transformation, excluding regions influenced by domain boundaries. Phase-contrast microscopy (Supplementary Fig. 2) was conducted using a Crystalline Tester® CP1 (Ceramic Forum Co. Ltd., Japan), equipped with a 20× objective lens (NA = 0.5) with a phase plate and a condenser lens (NA = 0.78) with a ring aperture. Illumination was provided by a 405 nm LED. The chip was mounted on a cover glass placed on the sample holder. The PCM image with a field of view of 359 × 300 $\mu m^2$ was acquired in 3 ms. PCM images were acquired by scanning the entire chip surface in the XY plane while focusing on the surface to confirm dislocation detection. PCM images were processed using a fast Image contrast was enhanced by applying a Fourier transform (FFT) bandpass filter (3 pixels < size < 40 pixels) using FIJI, an open-source image analysis platform built on ImageJ.[18]

Three-dimensional observation of dislocations using PCM was realized by shifting the focal plane of the objective lens in 2.5 μm increments from the surface toward the back surface of the chip, while maintaining the distance between the objective and condenser lenses. The actual focal displacement within the $\beta$-$Ga_2O_3$ crystal was calculated by multiplying the objective lens shift by the refractive index of $\beta$-$Ga_2O_3$ (n = 2).[19–21]

## III. RESULTS AND DISCUSSION

### A. Validation of Threading Dislocation Detection by PCM

Figure 1 shows (a) a PCM image and (b) an SR-XRT image of the same area ($0.67 \times 0.56$ mm$^2$). In the SR-XRT image, most observed features appear as bright dots corresponding to TDs, along with a few bright lines indicating dislocations parallel to the chip surface.[4,5,7,22–26] A total of 288 bright dots are visible in Fig. 1(b), of which 277 have positions that do not overlap with artifacts in the PCM image in Fig. 1(a). One example of such an artifact is the appearance of concentric rings, which can be caused by surface dust. A comparison was made between the contrasts in the PCM image and the positions of the bright dots in the SR-XRT image. The contrast marked by the yellow arrow represents a one-to-one correspondence between the PCM contrast and a bright dot in the SR-XRT image. In contrast, the pink arrow indicates a PCM contrast that is not visible at the surface but appears when the focus is adjusted to the interior, indicating a subsurface dislocation —defined here as a dislocation that does not emerge at the surface— which is also observed in SR-XRT. There are 10 bright dots in the SR-XRT image (indicated by brue arrows) with no corresponding contrast in the PCM image. Therefore, it was confirmed that 267 out of the 277 dislocations (yellow and pink arrows) observed in the SR-XRT image are also detected in PCM, resulting in a detection rate of 96.4% for threading dislocations. In contrast, 14 contrasts in the PCM image (indicated by blue arrows) have no matching bright dots in the SR-XRT image. In addition, no PCM contrasts corresponding to the bright lines indicated by orange lines drawn parallel to them in Fig. 1(b) were found in Fig. 1(a), indicating that PCM has not yet demonstrated the ability to detect dislocations running parallel to the chip surface.[27]

Future work will aim to modify the setup and optimize measurement conditions to facilitate the detection of such dislocations.

**B. Spatial Resolution of PCM for Closely Spaced Dislocations**

The number displayed next to the yellow arrow in the SR-XRT image in Fig. 1(b) indicates the number of PCM contrasts corresponding to the marked bright spot. Figure 2 presents enlarged PCM and SR-XRT images of the same region, allowing a direct comparison of spatial resolution between the two techniques. Multiple contrasts in the PCM image align with a single bright spot in the SR-XRT image. For instance, the bright dot labeled A in Fig. 2(b) corresponds to contrasts $A_1$ and $A_2$ in Fig. 2(a). Similarly, bright dots B, C, and D correspond to contrasts $B_1$ and $B_2$, $C_1$–$C_4$, and $D_1$–$D_3$, respectively. These results indicate that PCM can resolve closely spaced threading dislocations that appear as a single feature in SR-XRT.In SR-XRT, elastic strain fields surrounding dislocation cores are visualized as bright dots and the strain detection sensitivity is on the order of $10^{-7}$ δd/d. This high sensitivity causes the dislocation contrast to extend over several tens of micrometers. When dislocations are located within several micrometers of each other, these strain-contrast regions can overlap, making it difficult to distinguish them individually. Hence, the effective resolution of SR-XRT is limited by the dimensions of these bright dots, typically around 7 μm along the minor axis and 13 μm along the major axis, which are consistent with reported XRT values for wide-bandgap semiconductors[28,29].

In contrast, the PCM image clearly resolves $A_1$ and $A_2$ as separate features, even at a spacing of 6.5 μm, demonstrating PCM's ability to resolve in detecting closely spaced dislocations. PCM detects the phase retardation induced by dislocations; however, the region exceeding the detection sensitivity may be confined to the vicinity of the dislocation core, which could lead to more localized contrast and hence better in-plane resolution.

**C. Three-Dimensional Reconstruction of Threading Dislocations**

The reconstructed three-dimensional (3D) dislocation structure, derived from stacked and binarized PCM images using NIS-Elements software[30] (Nikon, Japan), is presented in Fig. 3. The depth range was constrained to 325 μm to minimize shadowing caused by scratches on the cover glass supporting the β-$Ga_2O_3$ chip. The dislocations are generally aligned parallel to the [010] direction, although local inclinations toward [001] and slight meandering on the (100) are evident. This 3D reconstruction enables tracking of dislocation propagation along the z-direction in β-$Ga_2O_3$.

**D. Slip System Analysis from Stacked PCM Images of Threading Dislocations**

Fig. 4(a) displays a projection image from the surface to a depth of 500 μm, reaching the back surface of the sample. This image was constructed in FIJI by selecting the minimum intensity value across the entire image stack at each XY pixel. Figs. 4(b)–(d) show magnified views of regions 1–3 in Fig. 4(a). Dark lines, representing the projections of dislocations, are visible in the image. Many of these lines align along the [001] direction, as illustrated by line A–A′ in Fig. 4(a), while a few align along [100], as shown by B–B′ in Fig. 4(b). The length and angle of each dark line were defined as the length and orientation (relative to the horizontal) of its longest axis after binarizing the PCM image. To selectively extract linear contrast features, binarized regions with a circularity ($(4\pi \times \text{area})/\text{perimeter}^2$) below 0.7 were retained, which effectively excluded shadows arising from overlapping lines or imaging artifacts. Fig. 5(a) presents a histogram of the in-plane angles of the dark lines observed in Fig. 4(a). The distribution peaks at 35–40°, which corresponds to the [001] direction (i.e., 36° relative to the horizontal axis of the chip). Fig. 5(b) shows a histogram of the lengths of the dark lines oriented within 36 ± 5° in the XY plane. The most frequent line length falls within the range of 5–10 μm. Considering the 500 μm stacking depth, this corresponds to an inclination angle of 0.6–1.1° from the surface normal. Most of these dislocations are regarded as nearly parallel to the [010] direction. Some dark lines reached lengths of up to 110 μm. Because the length of the dark line corresponds to the projected length of the dislocation line, this suggests that the dislocations are inclined by approximately 12.4° from the surface normal. It remains unclear whether this value reflects the maximum inclination detectable by PCM or the actual maximum inclination angle of dislocations in the observed $\beta$-$Ga_2O_3$ chip. Nonetheless, the presence of dislocations inclined toward the [001] direction clearly supports slip on (100) planes, as previously reported.[8,14,31] Likely slip systems include (100)<010>,[14] (100)<001>,[8,14] and (100)<0kl> (k, l ≠ 0). Projections of dislocation lines parallel to the [100] direction, inclined at 140° from the horizontal axis, are also detected, although they appear less frequently. Based on the histogram of dark lines within 140 ± 5° in the XY plane, the most common line length in this orientation is 0–5 μm, corresponding to an inclination of 0–0.6° from the surface normal. This observation indicates that most dislocations with projections parallel to [100] are also nearly parallel to [010]. Given that the [010] axis of the chip is inclined by ±0.2° toward a direction 35° clockwise from [102] within the (010) plane, it is reasonable to infer that the modal inclination angle of the dislocations is slightly greater toward [001] than toward [100]. Some lines extend up to 55 μm, corresponding to a 6.3° inclination, indicating that certain dislocations can slip on the (001) plane, as reported in Refs. 6, 8, 9, 13, 14, 32. Likely slip systems include (001)<010>,[8,9,13,14,32] (001)<100>,[6,9] and (001)<hk0> (h, k ≠ 0). A few dislocations with projections parallel to the [103] direction are also observed in Fig. 4(c), indicating the possibility of slip on the $(\bar{3}01)$ plane. In addition, winding lines are visible in Fig. 4(d). Since the movement of screw dislocations with a Burgers vector parallel to [010] is not restricted to a particular slip plane, such winding projections are plausible. These findings demonstrate that the slip planes and average inclination angles of dislocation lines relative to the surface normal can be deduced from the angles and lengths of their projections in the XY plane.

## IV. CONCLUSIONS

This study demonstrates that PCM can detect over 96 % of threading dislocations in $\beta$-$Ga_2O_3$ (010) observed by SR-XRT. The spatial resolution of PCM for closely spaced dislocations surpasses that of SR-XRT, and PCM can individually resolve dislocations separated by 6.5 μm. PCM facilitates the observation of the full three-dimensional structure of threading dislocations in a 10 × 15 × 0.5 mm $\beta$-$Ga_2O_3$ (010) chip by adjusting the focal plane and scanning the entire area. In addition, projecting the PCM image stack enables the tracing of dislocation propagation in the XY plane. The orientation and length of the traced dislocation lines reveal the slip planes and the inclination angles relative to the surface normal, respectively. As this method is applicable to light-transmitting single crystals, it provides considerable potential for nondestructive defect characterization in various wide-bandgap semiconductor wafers within a practically acceptable time frame for full-wafer measurement. Its capability to visualize the three-dimensional morphology and propagation of dislocations provides critical insights into dislocation behavior, making it a valuable technique for enhancing crystal quality and device performance in emerging material systems.

### ACKNOWLEDGMENT

This paper is based on results obtained from a project, JPNP22007, commissioned by the New Energy and Industrial Technology Development Organization (NEDO).

## AUTHOR DECLARATIONS

### Conflict of Interest

The authors have no conflicts to disclose.

### Ethics Approval

This study did not involve human participants or animal subjects and therefore did not require ethics approval.

### Author Contributions

**Yukari Ishiakwa**: Conceptualization (lead); Data curation (lead); Formal analysis (lead); Funding acquisition (equal); Investigation (equal); Writing – original draft (lead); Writing – review & editing (equal). **Daiki Katsube**: Investigation (equal); Writing – review & editing (equal). **Yongzhao Yao**: Investigation (equal); Writing – review & editing (equal). **Koji Sato**: Investigation (equal). **Kohei Sasaki**: Resources (lead); Funding acquisition (equal); Writing – review & editing (equal).

## DATA AVAILABILITY

The data that support the findings of this study are available within the article.

## REFERENCES

[1]K. Sasaki, Appl. Phys. Express **17**, 090101 (2024). https://doi.org/10.35848/1882-0786/ad6b73

[2]M. Higashiwaki, AAPPS Bull. **32**, 3 (2022). https://doi.org/10.1007/s43673-021-00033-0

[3]Y. He, F. Zhao, B. Huang, T. Zhang, and H. Zhu, Materials **17**, 1870 (2024). https://doi.org/10.3390/ma17081870

[4]H. Yamaguchi, A. Kuramata, and T. Masui, Superlattices Microstruct. **99**, 99 (2016); corrigendum **130**, 232 (2019). https://doi.org/10.1016/j.spmi.2016.04.030

[5]H. Yamaguchi and A. Kuramata, J. Appl. Cryst. **51**, 1372 (2018). https://doi.org/10.1107/S1600576718011093

[6]N. Mahadik, M. Tadjer, P. Bonanno, K. Hobart, R. Stahlbush, T. Anderson, and A. Kuramata, APL Mater. **7**, 022513 (2019). https://doi.org/10.1063/1.5051633

[7]S. Masuya, K. Sasaki, A. Kuramata, S. Yamakoshi, O. Ueda, and M. Kasu, Jpn. J. Appl. Phys. **58**, 055501 (2019). https://doi.org/10.7567/1347-4065/ab0dba

[8]Y. Yao, Y. Sugawara, and Y. Ishikawa, Jpn. J. Appl. Phys. **59**, 125501 (2020). https://doi.org/10.35848/1347-4065/abc1aa

[9]Y. Yao, Y. Sugawara, and Y. Ishikawa, J. Appl. Phys. **127**, 205110 (2020). https://doi.org/10.1063/5.0007229

[10]Y. Yao, Y. Sugawara, and Y. Ishikawa, Jpn. J. Appl. Phys. **59**, 045502 (2020).

https://doi.org/10.35848/1347-4065/ab7dda

[11]Y. Yao, Y. Ishikawa, and Y. Sugawara, J. Cryst. Growth **548**, 125825 (2020). https://doi.org/10.1016/j.jcrysgro.2020.125825

[12]Y. Yao, K. Hirano, Y. Sugawara, K. Sasaki, A. Kuramata, and Y. Ishikawa, APL Mater. **10**, 051101 (2022). https://doi.org/10.1063/5.0088701

[13]Y. Yao, Y. Tsusaka, K. Sasaki, A. Kuramata, Y. Sugawara, and Y. Ishikawa, Appl. Phys. Lett. **121**, 012105 (2022). https://doi.org/10.1063/5.0098942

[14]Y. Yao, Y. Tsusaka, K. Hirano, K. Sasaki, Y. Sugawara, and Y. Ishikawa, J. Appl. Phys. **134**, 155104 (2023). https://doi.org/10.1063/5.0169526

[15]R. Hattori, O. Oku, R. Sugie, K. Murakami, and M. Kuzuhara, Appl. Phys. Express **11**, 075501 (2018). https://doi.org/10.7567/APEX.11.075501

[16]R. Hattori, Y. Yao, and Y. Ishikawa, Mater. Sci. Forum, in press.

[17]A. Kuramata, K. Koshi, S. Watanabe, Y. Yamaoka, T. Masui, and S. Yamakoshi, Jpn. J. Appl. Phys. **55**, 1202A2 (2016). https://doi.org/10.7567/JJAP.55.1202A2

[18]J. Schindelin, I. Arganda-Carreras, E. Frise, V. Kaynig, M. Longair, T. Pietzsch, S. Preibisch, C. Rueden, S. Saalfeld, B. Schmid, and J. Y. Tinevez, Nat. Methods **9**, 676 (2012). https://doi.org/10.1038/nmeth.2019

[19]T. Onuma, S. Saito, K. Sasaki, T. Masui, T. Yamaguchi, T. Honda, A. Kuramata, and M. Higashiwaki, Jpn. J. Appl. Phys. **55**, 1202B2 (2016). https://doi.org/10.7567/JJAP.55.1202B2

[20]M. P. Anthony, *Electrical and Optical Investigations of Beta-Gallium Oxide*, Ph.D. thesis, California Institute of Technology, Pasadena (1972). https://thesis.library.caltech.edu/9645/1/Anthony_mp_1972.pdf

[21]Y. Bao, X. Wang, and S. Xu, J. Semicond. **43**, 062802 (2022).
https://doi.org/10.1088/1674-4926/43/6/062802

[22]T. Ohno, H. Yamaguchi, S. Kuroda, K. Kojima, T. Suzuki, and K. Arai, J. Cryst. Growth **260**, 209 (2004). https://doi.org/10.1016/j.jcrysgro.2003.08.065

[23]X. Zhang, M. Skowronski, K. X. Liu, R. E. Stahlbush, J. J. Sumakeris, M. J. Paisley, and M. J. O'Loughlin, J. Appl. Phys. **102**, 093520 (2007).
https://doi.org/10.1063/1.2809343

[24]W. M. Vetter, H. Tsuchida, I. Kamata, and M. Dudley, J. Appl. Cryst. **38**, 442 (2005).
https://doi.org/10.1107/S0021889805005819

[25]H. Yamaguchi, S. Nishizawa, W. Bahng, K. Fukuda, S. Yoshida, K. Arai, and Y. Takano, Mater. Sci. Eng. B **61–62**, 221 (1999). https://doi.org/10.1016/S0921-5107(98)00506-6

[26]Y. Yao, Y. Ishikawa, Y. Sugawara, Y. Takahashi, and K. Hirano, J. Electron. Mater. **47**, 5007 (2018). https://doi.org/10.1007/s11664-018-6252-3

[27]Y. Ishikawa, R. Hattori, and Y. Yao, in preparation.

[28]H.Yamaguchi, H. Matsuhata, and I. Nagai, Mat. Sci. Forum 600–603, 313(2009).
https://www.scientific.net/MSF.600-603.313

[29]H.Peng, T. Ailihumaer, Y. Liu, B. Raghotharmachar, X. Huang, L. Assoufid, M. Dudley, *J. Appl. Cryst.* **54**, 1225 (2021). https://doi.org/10.1107/S1600576721006592

[30]Nikon Instruments Inc., NIS-Elements,
https://www.microscope.healthcare.nikon.com/products/software/nis-elements

[31]S. Isaji, I. Maeda, N. Ogawa, R. Kosaka, N. Hasuike, T. Isshiki, K. Kobayashi, Y. Yao, and Y. Ishikawa, J. Electron. Mater. **52**, 5093 (2023). https://doi.org/10.1007/s11664-

023-10363-4

[32]H. Yamaguchi, S. Watanabe, Y. Yamaoka, K. Koshi, and A. Kuramata, Jpn. J. Appl. Phys. **61**, 045506 (2022). https://doi.org/10.35848/1347-4065/ac5adb

## FIGURES

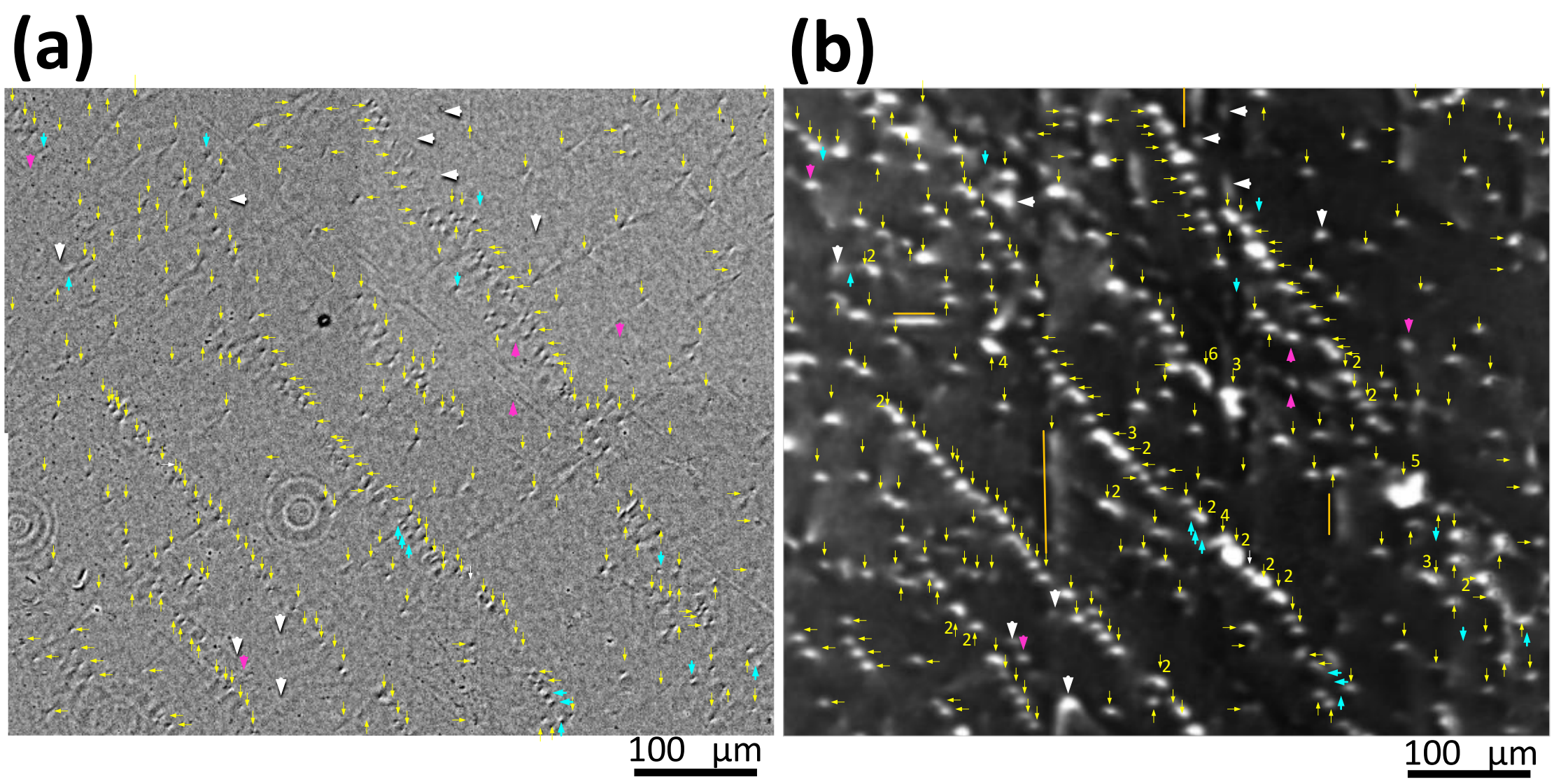

Fig. 1 (a) PCM image and (b) SR-XRT image of the same area. Yellow arrows indicate contrasts observed in both (a) and (b). Pink arrows denote features that appear in (a) only when the focal plane is shifted deeper, corresponding to the same positions as in (b). White arrows indicate bright contrasts visible only in (b), while blue arrows indicate contrasts observed exclusively in (a).

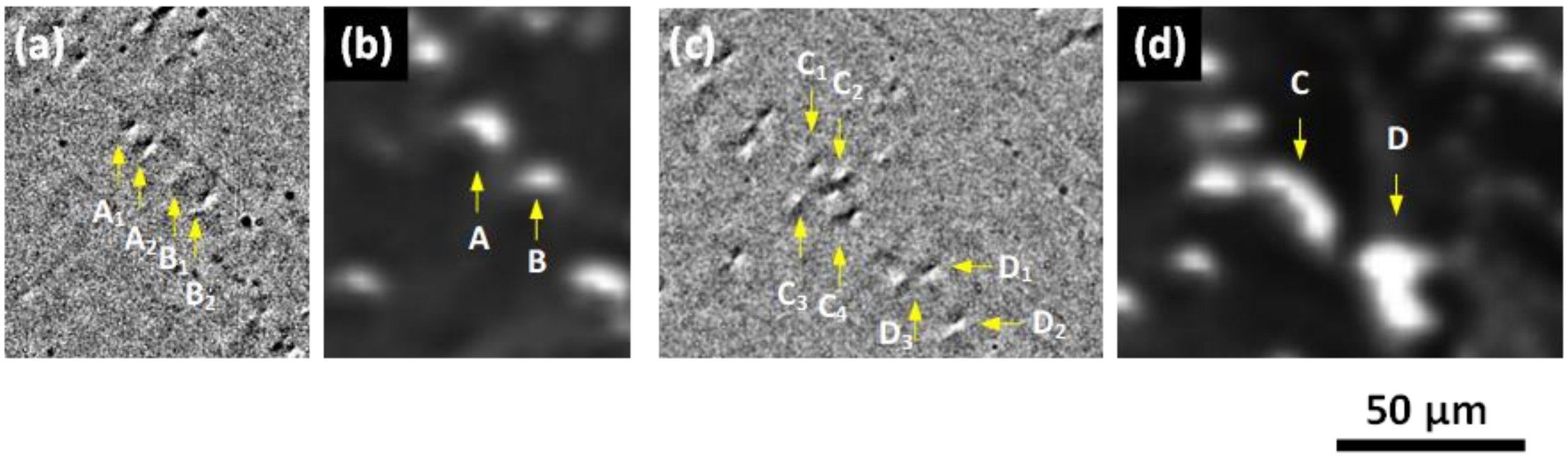

Fig. 2 PCM image (a) and SR-XRT image (b) were taken from the same region, and images (c) and (d) correspond to another region. In (a) and (c), contrasts $A_1$ and $A_2$ ($B_1$ and $B_2$) correspond to A (B). In (c) and (d), contrasts $C_1$–$C_4$ and $D_1$–$D_3$ correspond to C and D, respectively.

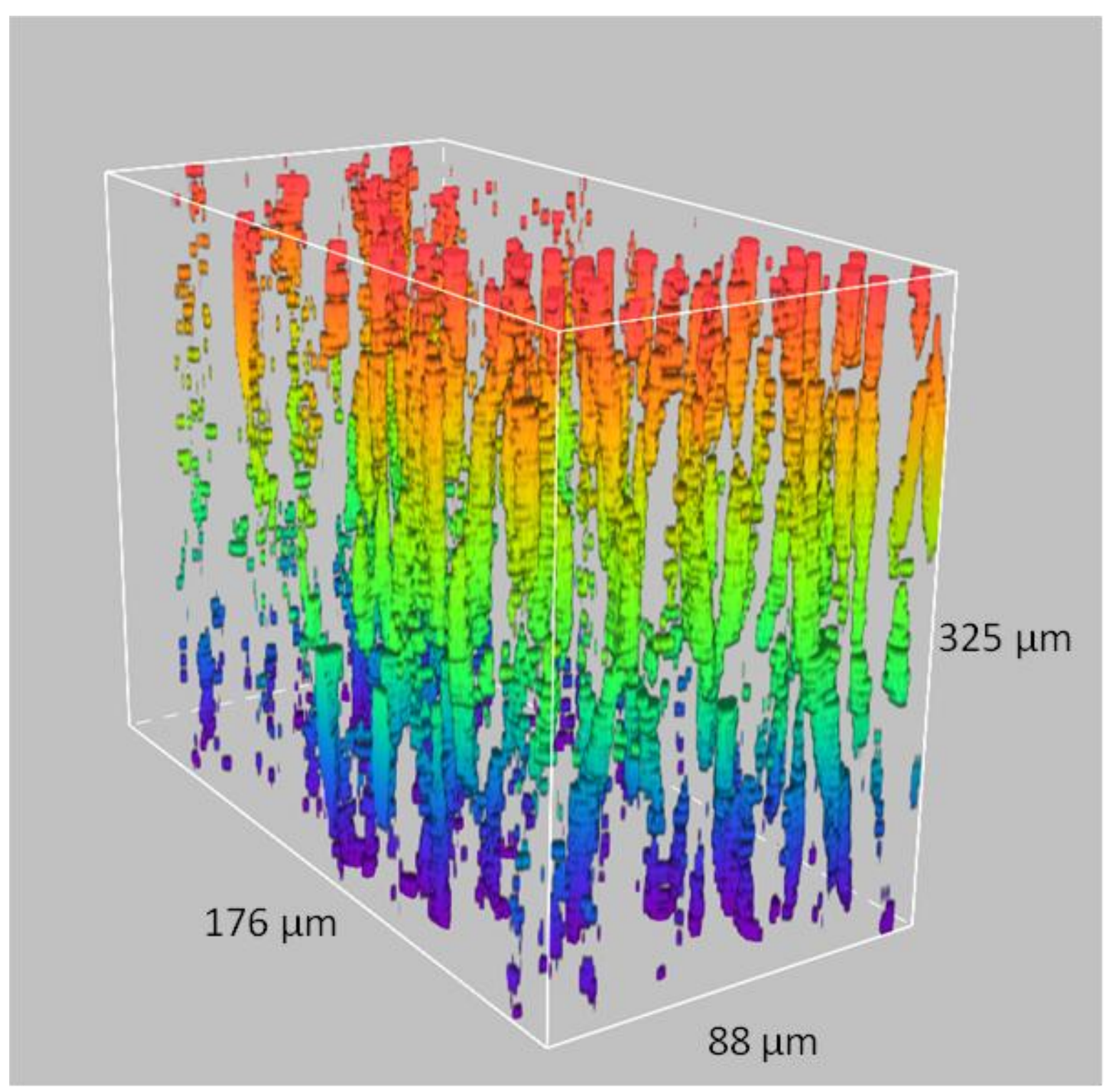

Fig. 3 Reconstructed three-dimensional dislocation structure within a volume of 176 μm × 88 μm × 325 μm, obtained from a stack of PCM images.

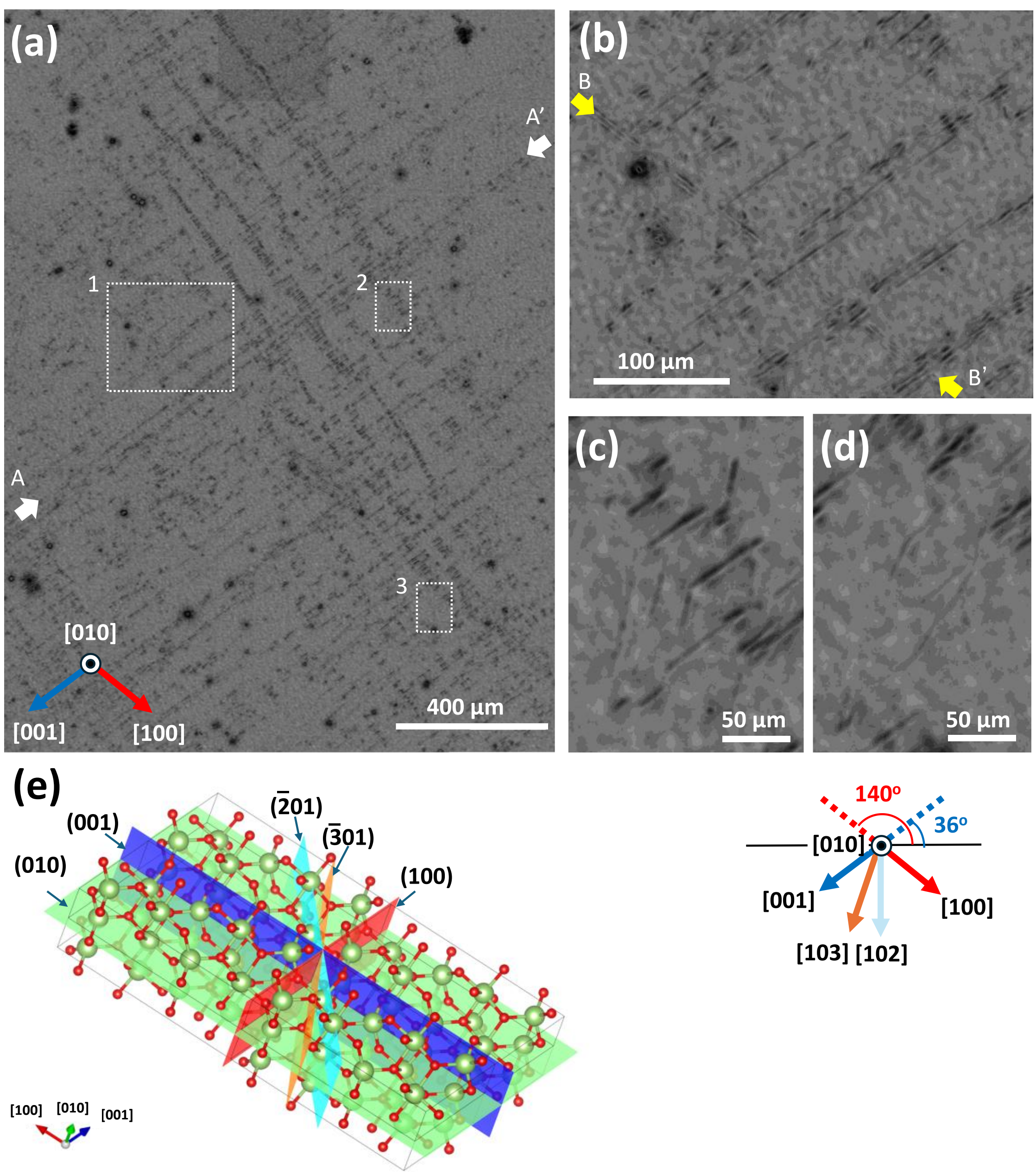

Fig. 4 (a) Projection image from the surface to the back surface, created by selecting the minimum intensity value at each pixel across the entire PCM image stack. (b)–(d) Enlarged views of regions 1–3 in (a). (e) Schematic of the crystal structure and representative crystallographic planes of $\beta$-$Ga_2O_3$.

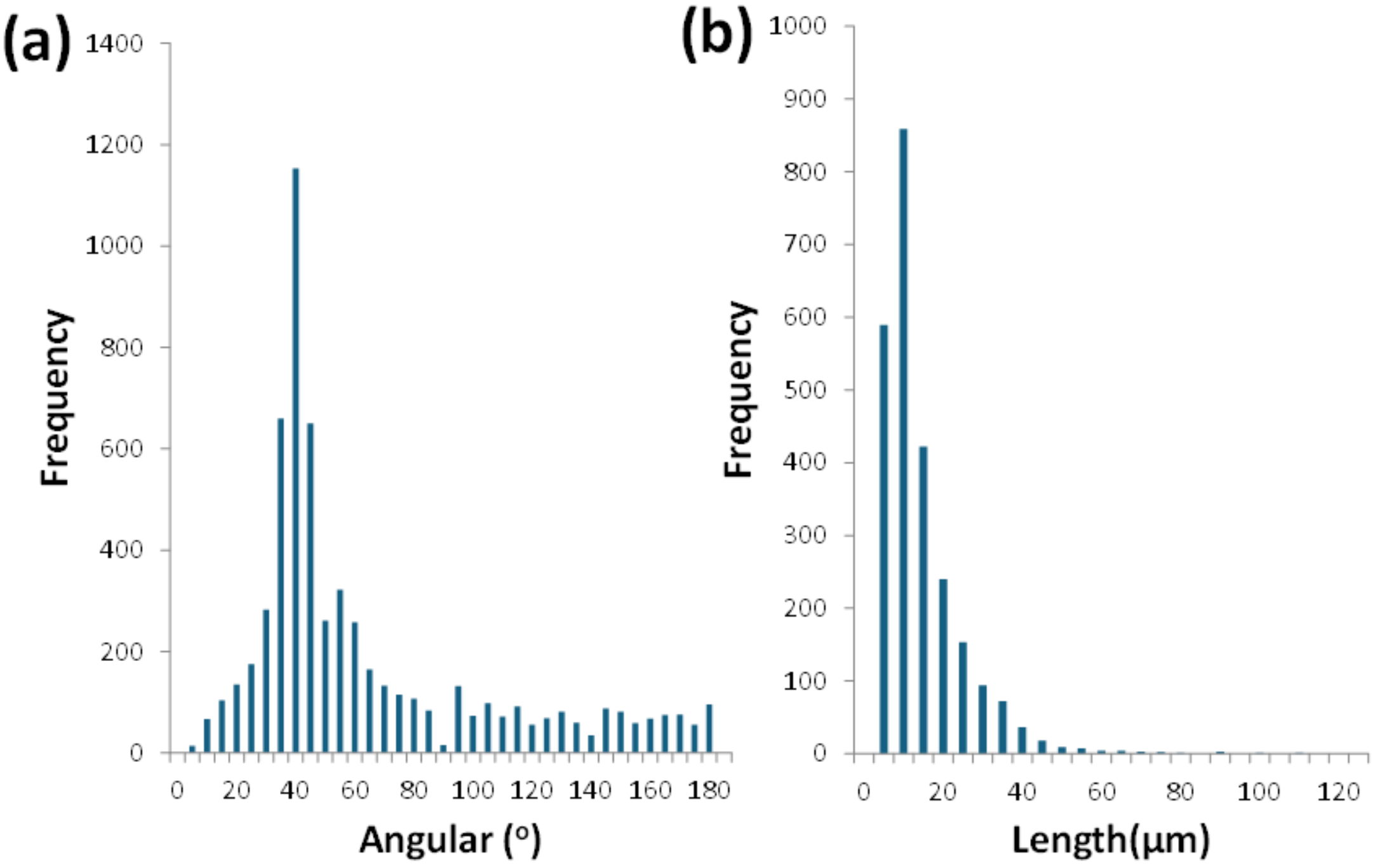

Fig. 5 (a) Histogram of the in-plane angles of the dark lines shown in Fig. 4(a).

(b) Histogram of the lengths of dark lines oriented within 30°–45° in the XY plane, corresponding to [001] direction.

The angle and length were defined as those of the major axis of binarized features with circularity < 0.7, following SR-XRT image processing.

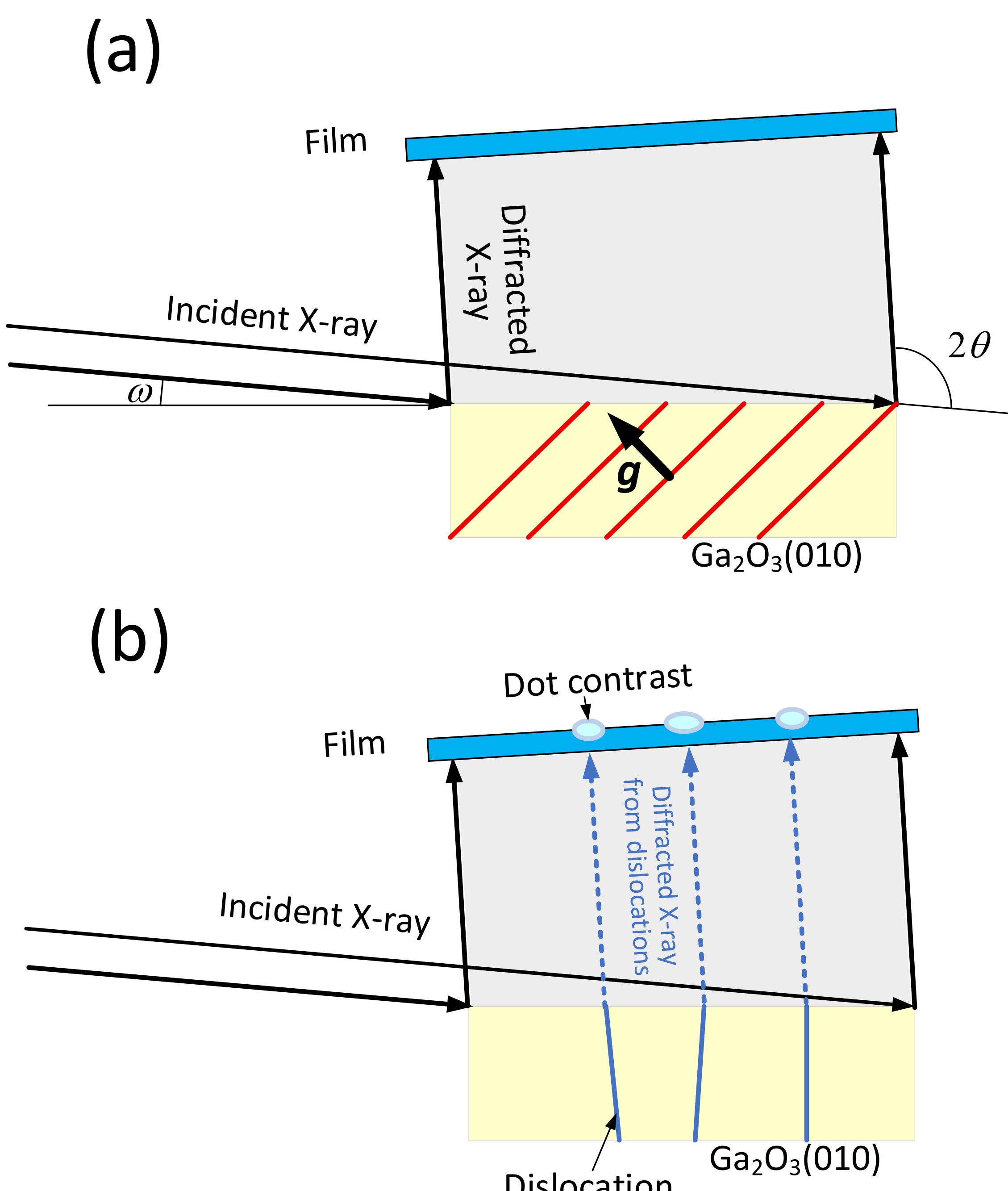

**Supplementary Fig. 1.** (a) Schematic illustration of the optical setup for X-ray topography. (b) Schematic diagram of dislocation imaging. Threading dislocations produce dot-like contrasts when the inner product of **g** and **b** is nonzero ($g \cdot b \neq 0$).

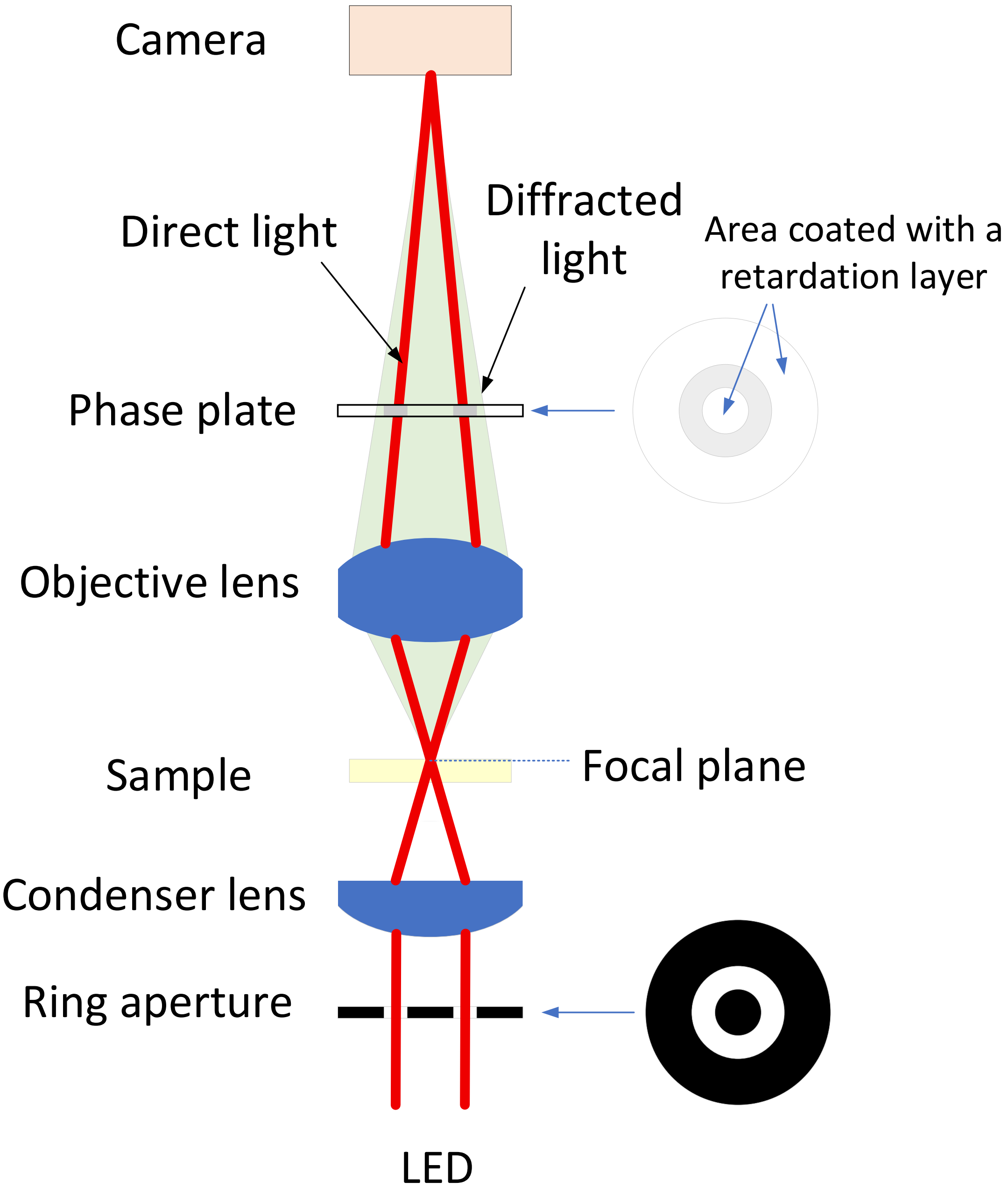

**Supplementary Fig. 2.** Schematic illustration of the optical setup of the phase-contrast microscope. The direct and diffracted lights interfere to form a phase-contrast image on the camera plane. Three-dimensional imaging was carried out by shifting the focal plane vertically with respect to the sample.